\documentclass[pra,aps,showpacs,twocolumn]{revtex4}
\usepackage{epsfig,amsmath,amssymb,amscd,latexsym,theorem,lscape,graphicx}
\newcommand{\ket}[1]{|#1\rangle}
\newcommand{\bra}[1]{\langle #1|}
\newcommand{\Tr}{\text{Tr}}
\begin{document}
\title{Geometry of decomposition dependent evolutions of mixed states}
\author{David Kult\footnote{Electronic address: 
david.kult@kvac.uu.se} and 
Erik Sj\"{o}qvist\footnote{Electronic address: erik.sjoqvist@kvac.uu.se}} 
\affiliation{$^{1}$Department of Quantum Chemistry, Uppsala University, 
Box 518, Se-751 20 Uppsala, Sweden.}
\date{\today}
\begin{abstract}
We examine evolutions where each component of a given decomposition of
a mixed quantal state evolves independently in a unitary fashion. The 
geometric phase and parallel transport conditions for this type of
decomposition dependent evolution are delineated. We compare this 
geometric phase with those previously defined for unitarily
evolving mixed states, and mixed state evolutions governed by
completely positive maps.
\end{abstract}
\maketitle 
\section{Introduction}
The concept of quantal geometric phase, first discovered 
for cyclic adiabatic evolutions by Berry \cite{berry84}, has 
been generalized in several steps. Aharonov and Anandan 
\cite{aharonov87} removed the restriction of adiabaticity and pointed 
out that the pure state geometric phase is due to the curvature of
projective Hilbert space. A general setting of the quantal geometric
phase including noncyclic evolution and sequential projection
measurements was put forward by Samuel and Bhandari \cite{samuel88},
based upon Pancharatnam's early work \cite{pancharatnam56} on
interference of classical light rays in distinct states of
polarization. Extension of the geometric phase towards the mixed state
case was first conceived by Uhlmann \cite{uhlmann86}, who introduced
parallel transport and concomitant geometric phase of density
operators.  Later, Sj\"{o}qvist {\it et al.} \cite{sjoqvist00}
discovered an operationally well-defined geometric phase for unitarily
evolving nondegenerate density operators in the context of quantum
interferometry. This latter phase concept has been generalized
\cite{filipp03,ericsson03a,singh03} and experimentally tested 
\cite{du03}. 

In brief, the mixed state geometric phase in Ref. 
\cite{sjoqvist00} is basically an extension of Pancharatnam's
relative phase between distinct physical states, added to it the
standard pure state parallel transport for each of the spectral basis
states of the density operator. Explicitly, for a mixed state,
initially described by the density operator $\rho (0)$, the quantity
\begin{equation}
\gamma = \arg \Tr \big( \rho (0) U (\tau) \big)
\label{eq:panch}
\end{equation} 
measures the Pancharatnam relative phase between the density operators
$\rho (0)$ and $\rho (\tau) = U(\tau) \rho (0) U^{\dagger} (\tau)$, 
$U(\tau)$ being unitary. Now, under the condition that $U(t)$, $t \in 
[0,\tau]$, continuously parallel transports the spectral basis of the 
density operator along some unitary path ${\cal C}$ ending at $\rho (\tau)$, 
the Pancharatnam relative phase in Eq. (\ref{eq:panch}) becomes the mixed 
state geometric phase $\gamma[{\cal C}]$ associated with this unitary path. 
Mathematically, $\gamma[{\cal C}]$ may be regarded the holonomy of 
a fiber bundle with structure group being the $N$ torus $T^N$. 

Many properties of mixed states may be understood in terms of
purifications of the considered system's density operator $\rho$,
i.e., by adding an ancilla system so that the whole system is in a
pure state whose partial trace over the ancilla is $\rho$. From this
perspective, the absence of unique concept of mixed state geometric
phase may be considered a consequence of the fact that nonpure states 
can be purified in many ways. While the above described approach in
Ref. \cite{sjoqvist00} arises naturally in standard one-particle
interferometry, Ref. \cite{uhlmann86} has been shown
\cite{ericsson03b} to depend also upon operations on the ancilla.

In this paper, we propose another form of geometric phase that has a
direct physical relation to the decomposition freedom of mixed states
\cite{peres95}, and which reduces to Ref. \cite{sjoqvist00} for
one-term decompositions. We shall see that the associated gauge
symmetry for such parallel transport has a fiber bundle interpretation
and that it plays a role in a certain type of evolution, for which
each component of the decomposition evolves independently in a unitary
fashion. This kind of evolution, which we shall call decomposition
dependent, is shown to have a natural interpretation in terms of
conditional unitary dynamics
\cite{barenco95} acting on separable mixed states in the space of the
considered system and some additional ancilla.

As a preliminary, we briefly describe, in the next section,
decompositions of density operators related to the freedom in the
preparation of mixed states. Decomposition dependent parallel
transport and mixed state geometric phase are introduced and analyzed
in Sec. III.  In Sec. IV, we examine particular instances of
decomposition dependent evolution related to the unitary case and
completely positive maps of the mixed state. The paper ends with the
conclusions.

\section{Decomposition Freedom}
Consider a preparation machine equipped with instructions to prepare 
a set of orthonormal pure states $\{ |k\rangle \}$, each member of 
which with probability $w_k$. The resulting mixed state may then be 
represented by the density operator
\begin{equation}
\rho = \sum_{k=1}^N w_k |k\rangle \langle k| ,  
\label{eq:decomp1}
\end{equation}
$N$ being the dimension of the considered system's Hilbert space 
${\cal H}$. Another, perhaps physically more feasible, preparation 
machine may instead prepare, with probabilities $\lambda_k$, the states 
$\rho_k$, yielding the density operator 
\begin{equation} 
\varrho = \sum_{k=1}^M \lambda_k \rho_k,
\label{eq:decomp2}
\end{equation}
where the states $\rho_1,\ldots,\rho_M$ may be nonpure. If $\varrho =
\rho$, then the two machines prepare the same mixed state, but in
different ways.  The different preparations of the same mixed state
are called decompositions, and one can show that all mixed states
(nonpure states) can be decomposed in infinitely many ways
\cite{peres95}.  The decomposition displayed by Eq. (\ref{eq:decomp1})
into the eigenbasis of the density operator is called the spectral
decomposition.

Given a mixed state represented by $\rho$ one cannot determine its
decomposition, since the outcome of any measurement only depends on
$\rho$, i.e., measurements are decomposition independent. Yet, we
claim there is a subtle difference between the decomposition of a
mixed state and the mixed state itself, because identifying the output
of a given preparation machine with the density operator representing
the state does not include any information about the details of the
preparation procedure, whereas identifying the output with the
corresponding decomposition does.

This latter observation makes it reasonable to introduce the set
${\cal S}_{\rho} = \{ \varrho | \varrho = \rho \}$ of 
all decompositions $\varrho$ of a density operator $\rho$ as being
an equivalence set with projection map $\Pi : {\cal S}_{\rho} 
\rightarrow \rho$. In the pure state case, each such equivalence set 
consist of a single element, as the decomposition is unique for such
states. On the other hand, for any nonpure state $\rho$, there are
infinitely many elements in ${\cal S}_{\rho}$. We may envisage paths
${\cal D}$ in the space ${\cal S} = \{ {\cal S}_{\rho} \}$ of all
decompositions of all mixed states. In the next section we discuss a
concept of geometric phase for such paths, but let us first examine
the space ${\cal S}$ a little bit more thoroughly.

Consider the set ${\cal A}$ of separable states of the form
\begin{equation}
\label{eq:decomp3}
\varrho_{sa}=\sum_{k=1}^M \lambda_k \rho_k \otimes \ket{\psi_k^a}
\bra{\psi_k^a},
\end{equation} 
where $\bra{\psi_k^a}\psi_l^a\rangle=\delta_{kl}$. These states act 
on Hilbert space ${\cal H} \otimes {\cal H}_a$, ${\cal H}$
being Hilbert space of the considered system and ${\cal H}_a$ is 
some $M$-dimensional ancillary space. Further, let us introduce the 
equivalence relation
\begin{equation}
\label{eq:decomp4}
\varrho_{sa} \sim (I \otimes U) \varrho_{sa} (I \otimes 
U^{\dagger}),  
\end{equation}
$U$ being unitary, and $I$ the identity operator. The set ${\cal A}/\sim$ 
of equivalent classes in 
${\cal A}$ under $\sim$ is isomorphic to the set ${\cal S}_M$ of 
decompositions into $M$ terms, where $M=\dim ({\cal H}_a)$.
Apparently ${\cal S}_M$ is a subset of ${\cal S}$, and in the following
we focus on this space, rather than on ${\cal S}$.  
Since ${\cal A}/\sim$ is isomorphic to the set ${\cal S}_M$ of $M$-term 
decompositions we may identify a decomposition of the form displayed
by Eq.  (\ref{eq:decomp2}) with a state of the form displayed by Eq.
(\ref{eq:decomp3}), keeping the equivalence relation in mind. 

\section{Geometry of Decomposition Dependent Evolutions} 
Let $U(t), t\in [0,\tau]$, be a continuous one-parameter family of
unitary operators with $U(0)=I$, $I$ being the identity operator on
${\cal H}$, and let $\{ \ket{k} \}$ be the eigenbasis of the initial
density operator $\rho (0)$, assumed to be nondegenerate
\cite{remark1}. Then, $U(t)$ is said to parallel transport the
density operator if it fulfills the conditions
\begin{equation}
\bra{k} U^{\dagger}(t)\dot{U}(t) \ket{k} = 0 , \ \forall k . 
\label{eq:ptm2}
\end{equation}  
Any parallel transporting operator is denoted by $U^{\parallel}(t)$ 
in the following. For such an operator, we may write the mixed state 
geometric phase associated with the path ${\cal C} : t\in [0,\tau] 
\rightarrow \rho (t) = U^{\parallel}(t) \rho (0) U^{\parallel 
\dagger}(t)$ as 
\begin{equation}
\gamma [{\cal C}] = \arg \Tr \big( \rho (0) U^{\parallel} (\tau) \big),  
\end{equation}
which is the total relative phase displayed by Eq. (\ref{eq:panch}) 
for parallel transported states. More generally, Singh {\it et al.} 
\cite{singh03} have put forward a kinematic approach, akin to that 
of Ref. \cite{mukunda93}, to the mixed state geometric phase in Ref. 
\cite{sjoqvist00}. They demonstrated that for any unitarity $U(t)$ 
the mixed state geometric phase reads 
\begin{equation}
\gamma [{\cal C}] = 
\arg \left( \sum_{l=1}^{N} w_l \bra{l} U(\tau) \ket{l} 
e^{-\int_0^{\tau}dt \bra{l} U^{\dagger}(t)\dot{U}(t) \ket{l}} \right).
\label{eq:ptm3}
\end{equation}
The generalization lies in the fact that the phase is geometric even 
if $U(t)$ does not fulfill Eq. (\ref{eq:ptm2}) and that it includes
the parallel transport condition, in the sense that the right-hand 
side of Eq. (\ref{eq:ptm3}) equals the total relative phase whenever
Eq. (\ref{eq:ptm2}) is fulfilled.

Let us now focus on evolutions of decompositions. Such evolutions may
be realized by starting from a state
\begin{equation}
\label{eq:larger1}
\varrho_{sa} (0) = \sum_{k=1}^M \lambda_k \rho_k \otimes 
|\psi_k^a\rangle\langle\psi_k^a|, 
\end{equation}
where $\langle\psi_k^a|\psi_l^a\rangle=\delta_{kl}$. Letting it
evolve under \cite{remark2}
\begin{equation}   
\label{eq:larger2}
U_{sa}(t) = \sum_{k=1}^M U_k(t) \otimes \ket{\psi_k^a} 
\bra{\psi_k^a}, 
\end{equation} 
where each $U_k(t), \ t\in [0,\tau]$, is a continuous one-parameter
family of unitarities with $U_k(0)=I$. Since $U_{sa}(t)$ keeps the
states $\ket{\psi_k^a}\bra{\psi_k^a}$ fixed, the evolution is well
defined in the space ${\cal A}/\sim$. Explicitly,  
\begin{eqnarray}
\label{eq:larger3}
\varrho_{sa}(t) & = &  U_{sa}(t) \rho_{sa} (0) 
U_{sa}^{\dagger}(t)  
\nonumber \\ 
 & = & \sum_{k=1}^M \lambda_k U_k(t) \rho_k U_k^{\dagger}(t) \otimes 
\ket{\psi_k^a} \bra{\psi_k^a},
\end{eqnarray}
which defines the decomposition dependent evolution 
\begin{equation}
\label{eq:larger4}
\varrho(t)=\sum_{k=1}^M \lambda_k U_k(t)\rho_kU_k^{\dagger}(t).
\end{equation}
One can realize this type of decomposition dependent evolution using
a type of preparation machine displayed by Fig. 1. Note that for $M=1$, 
we obtain ordinary unitary evolution of a mixed state, while for 
$M>1$ the evolution of the considered system is in general nonunitary. 

\begin{figure}[htb]
\centering
\includegraphics[width=8cm]{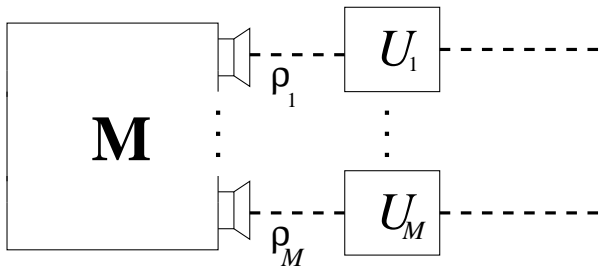}
\caption{Realization of decomposition dependent evolution. A 
machine ${\bf M}$ prepares outputs $\rho_k$, $k=1,\ldots,M$, each 
of which with probability $\lambda_k$ and conditionalized on the 
unitarity $U_k$.}
\end{figure}

By using an approach similar to that of Singh {\it et al.} 
\cite{singh03} we can define a geometric phase for this type of 
evolution. Let us first introduce a gauge transformation of the form
\begin{equation}
\widetilde{\varrho} (t) =  
\sum_{k=1}^{M} \widetilde{U}_k (t)\rho_k\widetilde{U}_k^{\dagger}(t),
\label{eq:ptd9}
\end{equation}
where
\begin{equation}
\widetilde{U}_k(t) = U_k (t)V_k(t),
\label{eq:ptd10}
\end{equation}
$V_k(t)$ being unitary and fulfilling $V_k(0)=I$. This transformation
may equivalently be expressed for the state $\varrho_{sa}(t)$ as
\begin{equation}
\label{eq:ptd101}
\widetilde{\varrho}_{sa}(t)=\widetilde{U}_{sa}(t)\varrho_{sa}(0)
\widetilde{U}_{sa}^{\dagger}(t), 
\end{equation} 
where 
\begin{eqnarray}
\label{eq:ptd102}
\widetilde{U}_{sa}(t) & = & U_{sa}(t) V_{sa}(t) , 
\nonumber \\ 
V_{sa}(t) & = & 
\sum_{k=1}^M V_k(t) \otimes \ket{\psi_k^a}\bra{\psi_k^a} , 
\end{eqnarray} 
$V_k(t)$ being unitary. The orbit of the decomposition remains unchanged 
under this transformation if $[\varrho_{sa}(0),V_{sa}(t)]=0$, which 
is fulfilled only if
\begin{equation}
\label{eq:ptd104}
[\rho_k,V_k(t)]=0, \ \forall k.
\end{equation}

Let us now introduce a concept of relative phase $\Gamma$ adapted 
to decomposition dependent evolutions. This is naturally given by 
considering the phase of the quantity $\Tr \left( \varrho_{sa} (0) 
U_{sa}(\tau) \right)$ yielding 
\begin{equation}
\label{eq:ptd11}
\Gamma = \arg \left( \sum_{k=1}^M \lambda_k 
\Tr \left(\rho_k U_k(\tau)\right)\right).  
\end{equation}
Note that for $U_k (\tau) = U(\tau), \ \forall k,$ this reduces 
to $\gamma$ in Eq. (\ref{eq:panch}). 
 
With the relative phase at hand we may now introduce the quantity
\begin{widetext}
\begin{equation}
\label{eq:ptd12}
\Gamma [{\cal D}] = \arg \left( \sum_{k=1}^M \lambda_k 
\sum_{l=1}^N w^k_l \langle k_l| U_k(\tau)|k_l\rangle 
e^{- \int_0^{\tau} dt \bra{k_l}U_k^{\dagger}(t)\dot{U}_k(t)\ket{k_l}} 
\right) ,  
\end{equation} 
\end{widetext}
where $\{ |k_l\rangle \}_{l=1}^N$ is the eigenbasis of $\rho_k$ and
$\{ w^k_l \}$ are the corresponding eigenvalues. $\Gamma [{\cal D}]$ is
gauge invariant since each term in the sum is invariant under the
corresponding transformation $U_k(t) \rightarrow \widetilde{U}_k(t)$
in Eq. (\ref{eq:ptd10}) fulfilling Eq. (\ref{eq:ptd104}). Moreover,
since $\Gamma [{\cal D}]$ is real-valued and reparametrization
invariant, we may define it as the geometric phase for decomposition
dependent evolutions. Demanding that the geometric phase equals the
total phase for parallel transported states provides us the following
$N \times M$ parallel transport conditions
\begin{equation}
\bra{k_l} U_k^{\parallel\dagger} (t) \dot{U}_k^{\parallel} (t) 
\ket{k_l} = 0, \ \forall k,l. 
\label{eq:ptd14}
\end{equation}
Apparently a given decomposition is parallel transported if each
component $\rho_k$ of the decomposition fulfills the parallel
transport conditions given in Ref. \cite{sjoqvist00}, with respect to
the unitary operator $U_k(t)$ acting on that specific component.

The geometric phase $\Gamma [{\cal D}]$ depends only on the path
${\cal D}$ in the space ${\cal S}_M$ of decompositions. Such a path
may be lifted to that of a pure state $\ket{\Psi (t)} \in {\cal H}
\otimes {\cal H}_a \otimes {\cal H}_b$ by attaching yet another
ancilla such that the projection map $\pi : \ket{\Psi (t)}
\rightarrow \Tr_b \ket{\Psi (t)} \bra{\Psi (t)}$ is ${\cal D}$. 
This is fulfilled by  
\begin{eqnarray} 
\ket{\Psi(0)} = 
\sum_{k=1}^M \sum_{l=1}^N \sqrt{\lambda_k} \sqrt{w_l^k} 
\ket{k_l} \otimes \ket{\psi_k^a} \otimes \ket{\psi_{kl}^b}, 
\end{eqnarray} 
evolving as $\ket{\Psi(t)} = U_{sab}(t) \ket{\Psi(0)}$, where
\begin{equation}
\label{eq:ptd16}
U_{sab}(t)= \sum_{k=1}^M U_k(t)  \otimes \ket{\psi_k^a} 
\bra{\psi_k^a} \otimes I^b.
\end{equation}
From this point of view, $\Gamma [{\cal D}]$ is the holonomy of 
the fiber bundle $({\cal S},{\cal H} \otimes {\cal H}_a \otimes 
{\cal H}_b, \pi,{\cal G})$ with structure group ${\cal G}$ being 
the $N \times M$ torus $T^{N \times M}$. 

\section{Relation to Evolutions of Mixed States}
Consider the map
\begin{equation}
\label{eq:map}
{\cal F}: \ \varrho_{sa} \longrightarrow \Tr_a \varrho_{sa},
\end{equation}
from the space ${\cal A}/\sim$ to the space of mixed states. Using
this map, and the fact that the space of decompositions is isomorphic
to ${\cal A}/\sim$, one can take a path ${\cal D}$ from the space of
decompositions into a path ${\cal C}$ in the space of mixed states. In
this section we consider special cases of decomposition dependent
evolutions producing paths ${\cal D}$, which corresponds to continuous 
sets of mappings ${\cal C}$ in the space of mixed states.

Let us first consider unitary evolutions of mixed states, 
which corresponds to the special case of decomposition dependent evolutions 
where $U_k(t)=U(t)V_k(t)$ and $[V_k(t),\rho_k]=0$. In this case, 
the geometric phase $\Gamma [{\cal D}]$ in Eq. (\ref{eq:ptd12}) 
becomes
\begin{eqnarray}
\label{eq:ptd15}
\Gamma [{\cal D}] & = & 
\arg \Big( \sum_{k=1}^M \lambda_k \sum_{l=1}^N w^k_l 
\langle k_l| U(\tau)|k_l\rangle 
\nonumber \\ 
 & & \times e^{-\int_0^{\tau} dt 
\bra{k_l} U^{\dagger}(t) \dot{U}(t) \ket{k_l}} \Big) , 
\end{eqnarray}  
which may in general be different from the geometric phase $\gamma 
[{\cal C}]$ introduced in Ref. \cite{sjoqvist00}. Hence, the additional
information about the decomposition affects the geometric phase, as
one could suspect. However, there exists a special case when the two
geometric phases numerically coincide, namely when all terms in the
decomposition have the same eigenbasis, i.e., $\{ \ket{k_l} \equiv 
\ket{l} \}$. Consequently, we obtain 
\begin{eqnarray}
\Gamma [{\cal D}] & = & 
\arg \Big( \sum_{k=1}^M \lambda_k \sum_{l=1}^N 
w^k_l \bra{l} U(\tau) \ket{l} 
\nonumber \\ 
 & & \times e^{-\int_0^{\tau} dt 
\bra{l} U^{\dagger} (t) \dot{U} (t) \ket{l}} \Big) ,
\end{eqnarray} 
i.e., $\Gamma [{\cal D}] = \gamma [{\cal C}]$ in Eq. (\ref{eq:ptm3}) 
by the identification $w_l = \sum_k \lambda_k w_l^k$. This may be 
regarded a consequence of the fact that one can choose $V_1 (t) = 
\ldots = V_M (t) \equiv V(t)$ in this case, which precisely corresponds 
to the gauge symmetry $T^N$ of the mixed state. On the other hand, if 
at least two of the $\rho_k$'s do not diagonalize in the same basis, 
the gauge group cannot be reduced to $T^N$ and $\Gamma [{\cal D}]$ 
cannot be associated with the holonomy of the unitarily evolving 
mixed state.  

One can further analyze the relation between decomposition dependent
evolution and unitary mixed state evolution by considering the latter
as decomposition dependent, where the decomposition contains a single 
term, i.e., $\rho=\rho$. This entails that the path ${\cal C} : t 
\rightarrow U(t) \rho U^{\dagger}(t)$ in the space of mixed states 
corresponds to the path $\widetilde{{\cal C}} : t \rightarrow
 U(t)\rho U^{\dagger}(t)\otimes \ket{\psi_1^a}
\bra{\psi_1^a}$ in the space ${\cal A}/\sim$.  This seems to be a
natural correspondence, and it is apparent that we have
$\Gamma[\widetilde{{\cal C}}]=\gamma [{\cal C}]$.  However, the path
$\widetilde{{\cal C}}$ can only be the same as a path ${\cal D}$
produced by a one-term decomposition. To see this, let us assume the
opposite, i.e.,paths of the form ${\cal D} : t \rightarrow
\sum_{k=1}^M \lambda_k U (t) \rho_k U^{\dagger}(t)\otimes
\ket{\psi_k^a} \bra{\psi_k^a}\}$. Then ${\cal D}$ correspond to $
\widetilde{{\cal C}}$ only if there exists a continuous one-parameter 
family of unitarities $V(t)$, such that
\begin{eqnarray}
\label{eq:ptd18}
& & \sum_{k=1}^M \lambda_k U (t)\rho_k U^{\dagger}(t)\otimes 
\ket{\psi_k^a} \bra{\psi_k^a}  =  
\nonumber \\
& & U(t)\otimes V(t) \, \rho\otimes\ket{\psi_1^a} \bra{\psi_1^a}
\, U^{\dagger}(t) \otimes V^{\dagger}(t) .
\end{eqnarray}
Tracing over the system part on both sides gives us
\begin{equation}
\label{eq:ptd19}
\sum_{k=1}^M \lambda_k \ket{\psi_k^a} \bra{\psi_k^a} = 
V(t)\ket{\psi_1^a} \bra{\psi_1^a} V^{\dagger}(t),
\end{equation}
where the right-hand side is a pure state, whereas the left-hand side
is a pure state only if all but one of the $\lambda_k$'s
vanishes. Hence, the previously discussed numerical agreement between
the geometric phase for a class of $M$-term decompositions and the
mixed state geometric phase, cannot be explained as a correspondence
between paths. Rather we have a situation where $\Gamma[{\cal D}] =
\Gamma [\widetilde{{\cal C}}]= \gamma [{\cal C}]$, ${\cal D}$ and
$\widetilde{{\cal C}}$ being distinct paths.

Another special case of decomposition dependent evolutions is  
\begin{equation}
\label{eq:cpm1}
\varrho(t)=\sum_{k=1}^M \lambda_kU_k(t)\varrho (0) U_k^{\dagger}(t),
\end{equation}
i.e., when $\rho_1 = \ldots = \rho_M = \varrho (0)$. The corresponding 
state $\varrho_{sa}(0)$ takes the product form
\begin{equation}
\varrho_{sa}(0) = 
\varrho(0) \otimes \left( \sum_{k=1}^M \lambda_k |\psi_k^a
\rangle\langle \psi_k^a|\right).
\end{equation} 
Being a product state implies that this evolution corresponds to mixed
state evolution governed by a continuous one-parameter family of
completely positive (CP) maps \cite{preskill}. This can also be seen
by reexpressing Eq. (\ref{eq:cpm1}) as
\begin{equation}
\label{eq:cpm2}
\varrho(t)=\sum_{k=1}^M W_k(t)\varrho (0) W^{\dagger}_k(t),
\end{equation}
which, for the mixed state, is the Kraus representation of a CP map,
where $W_k(t) = \sqrt{\lambda_k}U_k(t)$, and $\sum_{k=1}^M
W_k^{\dagger}(t)W_k(t)=I$. Not all CP maps have a Kraus
representation where $W_k(t)=\sqrt{\lambda_k}U_k(t)$, $U_k(t)$ being
unitary; only a fraction of all CP maps can be viewed as decomposition
dependent evolutions \cite{remark3}, but for this class of maps we may
define the geometric phase as
\begin{eqnarray}
\label{eq:cpm3}
\Gamma [{\cal D}] & = & 
\arg \Big( \sum_{k=1}^M \lambda_k \sum_{l=1}^N 
w_l \langle l| U_k(\tau)|l\rangle 
\nonumber \\ 
 & & \times e^{- \int_0^{\tau} dt 
\langle l |U_k^{\dagger}(t)\dot{U}_k(t)|l\rangle} \Big),
\end{eqnarray}   
where $\{|l\rangle\}$ is the eigenbasis of $\varrho (0)$, and $w_l$
are the corresponding eigenvalues. The concomitant parallel transport
conditions reads
\begin{equation}
\label{eq:cpm4}
\langle l |U_k^{\dagger}(t)\dot{U}_k(t)|l\rangle=0, \ \forall k,l.
\end{equation}   
 
The geometric phase for CP maps has previously been considered 
in Ref. \cite{ericsson03a}. In the special case of Kraus operators 
of the form $W_k (t) = \sqrt{\lambda_k} U_k(t)$, $k=1,\ldots,M$, 
this approach associates, at each $t$, a relative phase 
$\Gamma_k (t)$ to each $W_k (t)$ by the expression  
\begin{equation} 
\label{eq:cpm41}
\nu_k e^{i\Gamma_k (t)} = 
\sqrt{\lambda_k} \Tr \left[ \varrho (0) U_k (t) \right] . 
\end{equation}
By introducing parallel transport conditions in Eq. (\ref{eq:cpm4}), 
the $M$ geometric phases are given by
\begin{eqnarray}
\label{eq:cpm5}
\widetilde{\Gamma}_g^{(k)} = 
\arg \left(\sum_{l=1}^N w_l \langle l| 
U_k(\tau)|l\rangle e^{- \int_0^{\tau} dt \bra{l} 
U_k^{\dagger}(t)\dot{U}_k(t) \ket{l}} \right) .  
\nonumber \\  
\end{eqnarray}
It follows that $\Gamma_g$ and $\widetilde{\Gamma}_g^{(k)}$ are related as
\begin{equation}
\label{eq:cpm7}
\Gamma [{\cal D}] = \arg \left( \sum_{k=1}^M 
\lambda_k r_k e^{i\widetilde{\Gamma}_g^{(k)}}
\right) , 
\end{equation}
where 
\begin{equation}
\label{eq:cpm6}
r_k \equiv \left| \sum_{l=1}^N w_l 
\bra{l} U_k(\tau) \ket{l} e^{-\int_0^{\tau} dt \bra{l} 
U_k^{\dagger}(t)\dot{U}_k(t) \ket{l}} \right| ,  
\end{equation}
is the mixed state visibility \cite{sjoqvist00} for the unitary evolution 
$\varrho (0) \rightarrow U_k(t) \varrho (0) U_k^{\dagger} (t)$. 

\section{Conclusions} 
We have introduced a concept of decomposition dependent evolution of
quantal states and discussed the concomitant geometric phase and
parallel transport. This geometric phase depends only upon the path in
the space of all decompositions and is different, both conceptually
and numerically, from the geometric phase of mixed states. It may even
differ from the standard geometric phase for mixed states
\cite{sjoqvist00} in the case of unitary evolution of the decomposition. 
We have further demonstrated that the concept of geometric
phase for decompositions and that of the corresponding mixed state in
the unitary case, become identical if each component of the
decomposition diagonalize in the same basis. We have also shown 
that the present approach leads to a notion of geometric phase for a
special class of completely positive maps that essentially differs
from previous suggestions
\cite{ericsson03a,peixoto03}.
\section*{Acknowledgments}
We wish to thank Johan {\AA}berg for useful discussions. The work by
E.S. was supported by the Swedish Research Council.
 

\begin{thebibliography}{99}
\bibitem{remark2} More general types of conditional dynamics of 
the form $\Phi_{sa,t} = \sum_{k=1}^M \Phi_{k,t} \otimes \ket{\psi_k^a} 
\bra{\psi_k^a}$, where each $\rho \rightarrow \Phi_{k,t} [\rho]$ is 
a continuous one-parameter family of completely positive maps, can 
be envisaged, but will be treated elsewhere.
\end{thebibliography}
\end{document}